# Universal Control of Symmetric States Using Spin Squeezing

**Nir Gutman**[1,†]**, Alexey Gorlach**[1,†]**, Offek Tziperman**[1]**, Ron Ruimy**[1] **and Ido Kaminer**[1,*]

[1]*Technion – Israel Institute of Technology, Haifa 32000, Israel*

[†]*equal contributors,* [*]*kaminer@technion.ac.il*

**The manipulation of quantum many-body systems is a frontier challenge in quantum science. Entangled quantum states that are symmetric to permutation between qubits are of growing interest. Yet, the creation and control of symmetric states has remained a challenge. Here, we find the way to universally control symmetric states, proposing a scheme that relies solely on coherent rotations and spin squeezing. We present protocols for the creation of different symmetric states including Schrödinger's cat and Gottesman-Kitaev-Preskill states. The obtained symmetric states can be transferred to traveling photonic states via spontaneous emission, providing a powerful mechanism for the creation of desired quantum light states.**

# Control of symmetric-state systems and their applications

Universal control of quantum systems is a critical building block in quantum computing and more generally in quantum information processing. In conventional qubit-based quantum systems, single-qubit gates and a two-qubit entangling gate provide all the ingredients to obtain universal control over an arbitrary number of qubits [1]. Despite this, the creation of most states requires the sequential application of a number of gates that grows exponentially with the number of qubits [2]. This becomes impractical for most states, even for a modest number of qubits (e.g., 40 qubits). A large number of gates is incompatible with current architectures of quantum computers that suffer from short coherent times [3]. Hence, sets of operations that can create desired quantum states with only polynomial number of gates are highly sought after.

Recent years have shown increasing interest in quantum states having symmetry to permutation between any two constituent qubits. These states that are used in many platforms, such as ones encoding the quantum information in harmonic oscillators (continuous variables), in multi-level systems (qudits), and in ensembles of indistinguishable quantum particles.

Famous systems in which symmetric states naturally arise include nitrogen-vacancy centers [4], nuclear magnetic resonance systems [5], superconducting circuits [6,7], trapped ions [8,9], neutral atoms [10,11], and quantum dots [12,13]. The individual quantum particles comprising these systems are different, yet the combined system is described by the same Hilbert space of symmetric states. For an operation to keep the state inside the symmetric Hilbert space, it should act symmetrically on the entire population of particles. The simplest such operations are coherent rotations acting simultaneously on the state of every particle and spin squeezing of the combined state of all particles. These operations have been thoroughly explored in theory [14,15] and demonstrated in experiments [16–20], for example, enabling the creation of atomic Schrödinger's cat states [21,22]. Nevertheless, it remained unknown whether such operations can create any arbitrary symmetric state and fully control the symmetric system.

Here we show that spin squeezing together with coherent rotations constitute universal control over symmetric states. Any arbitrary symmetric state can be created using a polynomial number of symmetric operations, rather than the exponential number of operations needed in existing conventional approaches. Our findings directly apply to any physical system described by symmetric states that allow coherent rotations and squeezing.

Coherent rotations have been implemented in numerous systems [18,19,23,24], but they are not sufficient for universal control. To achieve universality, we find it sufficient to add the spin-squeezing operation. Spin squeezing is a ubiquitous operation that acts not only on spin systems, but on any symmetric system, such as atoms, quantum dots or superconducting circuits [21,24–28]. For example, spin squeezing was implemented on atoms in a cavity using multiple lasers with different frequencies [16,28–30], or by interaction with off-resonant light [31]. Innovative implementations of coherent control and spin squeezing in the last decade [26–28] have made our proposal for universal control relevant for all state-of-the-art platforms in quantum science.

**Direct transfer of symmetric states to photonic states**

In cases where the symmetric system describes quantum emitters, our findings have another intriguing implication. We rely on a unique property of symmetric states: when undergoing spontaneous emission, symmetric-state superpositions are precisely the states that get transferred onto a pulse of light in the form of a *single* quantum-optical mode [32]. This situation is unique because in many other cases, the light emitted in spontaneous emission occupies many spatiotemporal modes, and observing only part of them is expected to create a decohered mixed state. i.e., the initial quantum state of the emitters will not be preserved due to the inherent "random" nature of spontaneous emission.

In the language of quantum information, spontaneous emission can be considered an error channel that destroys the information embedded in the qubits. However, in our case spontaneous emission is a promising way to transfer pure state of emitters (static qubits) to almost single mode pure state of light (flying qubits) that can be used to store and transmit quantum information with minimal loss.

Using this direct transfer of symmetric states to light pulses, the universal control of symmetric states allows us to create high-fidelity quantum light states that are desired for quantum information processing. Particularly, the fields of photonic quantum computation and communication are in search for an efficient way of creating specific quantum states of light, such as the Schrödinger cat and Gottesman-Kitaev-Preskill (GKP) states [33]. These light states are necessary resources for continuous-variable quantum information processing [34], a rising approach that complements the conventional, discrete-variable, qubit encoding.

To find efficient protocols for creating any arbitrary photonic state and specifically the ones desired for quantum information processing, we develop an optimization protocol for controllable generation of arbitrary symmetric states. We find sequences of squeezing and rotations that create photonic square and hexagonal GKP states of 10 dB squeezing with 95% and 94% fidelity, as well as two- and four-legged Schrödinger's cat states with 97% and 94% fidelity, respectively. Our results convey the importance of studying new methods for manipulating systems of quantum emitters, especially for the ambitious goal of achieving universal control of light in various spectral regimes.

**Creation of quantum light using systems of emitters**

Gaussian states of harmonic oscillators, i.e., states that have Gaussian Wigner functions [35], such as squeezed light, can be beneficial for sensing in spectroscopy and metrology [36]. Creation of these states requires coherent displacements and squeezing. In contrast, non-Gaussian states of light such as Schrödinger's cat and GKP states, cannot be created using only those two operations [35]. Creating non-Gaussian states in conventional photonic systems can only be done by acquiring strong higher-order nonlinearities such as Kerr nonlinearity [35,37] or conditional operators and post-selections [38–41]. However, these nonlinearities are relatively weak and inefficient in the optical range. Therefore, the field of quantum optics is in search for new ways of creating non-Gaussian quantum states of light.

The approach that we explore here for generating quantum light uses systems of emitters, where the emitters themselves are the nonlinear element. Specifically, quantum emitters can generate non-Gaussian light states through their spontaneous emission to waveguides or optical cavities [32]. The simplest example of a non-Gaussian state is the single-photon Fock state, created regularly by spontaneous emission from a single emitter. Similar approaches that rely on spontaneous emission are implemented, for example, in superconducting qubits [40,42], in strongly interacting Rydberg atoms [43,44], and in other atoms coupled to optical cavities [45] or to nanophotonic waveguides [37,38]. Despite the success in particular cases of generation of non-Gaussian states, it was unknown to what extent the emitters can be manipulated and what kinds of quantum light states can be created this way.

Our work shows that the lowest possible non-linear operations on the emitters – coherent rotations and spin squeezing – already enable the creation of arbitrary symmetric states (Fig. 1). Then, spontaneous emission transfers the quantum state, enabling controllable generation of arbitrary photonic states [32,46,47].

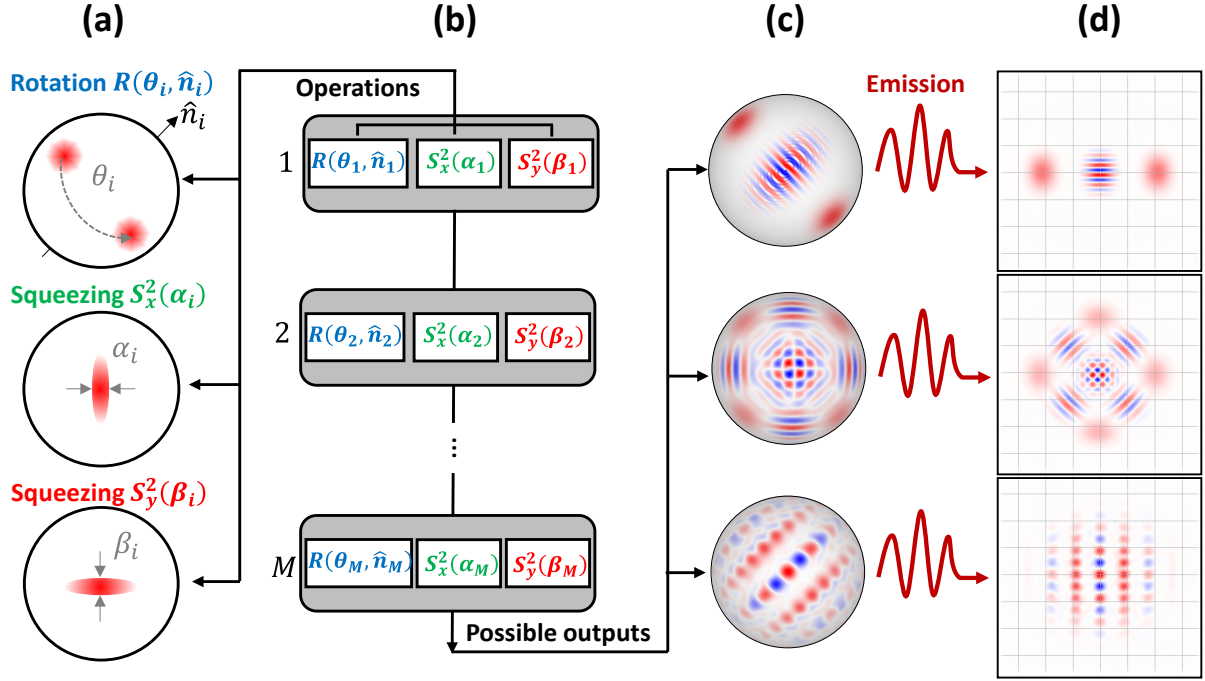

Figure 1: **Universal control of quantum states using spin squeezing and rotations**. (a,b) Sequences of the three operations of coherent rotation $R(\theta_i, \hat{n}_i)$ and squeezing with $S_x^2(\alpha_i)$ and $S_y^2(\beta_i)$ are sufficient for efficiently creating any arbitrary state in the Hilbert space of the permutationally symmetric states. (c) Example Wigner functions of the created states can be plotted on Bloch spheres [48], each representing the joint state of indistinguishable emitters. (d) The spontaneous emission by such emitters can be tailored to desired states of light, as illustrated by the corresponding Wigner functions of the emitted light pulses.

## Theory of symmetric states

We consider a system of $N$ two-level non-interacting emitters. We use the term emitters to describe general systems such as quantum dots, spins, transmons, trapped ions, etc. Each of these systems can be described in terms of symmetric states [49,50] that are invariant under any permutation of emitters. The basis of the symmetric states, sometimes called the Dicke ladder [50], is defined by:

$$\begin{cases} |N\rangle = |ee \dots e\rangle \\ |N-1\rangle = 1/\sqrt{N}(|ge \dots e\rangle + |eg \dots e\rangle + \cdots |ee \dots g\rangle) \\ \dots \\ |1\rangle = 1/\sqrt{N}(|eg \dots g\rangle + |ge \dots g\rangle + \cdots |gg \dots e\rangle) \\ |0\rangle = |gg \dots g\rangle \end{cases}.$$

Here $|g\rangle, |e\rangle$ are the ground and excited states of each two-level emitter. Our treatment is valid regardless of their energy gap (and consequently the frequency of the emitted light). We denote the symmetric operators $S_x = \sum_i \sigma_x^{(i)}$, $S_y = \sum_i \sigma_y^{(i)}$, $S_z = \sum_i \sigma_z^{(i)}$, where $\sigma_{x,y,z}^{(i)}$ are the Pauli matrices on the $i^{th}$ emitter. Arbitrary functions of these operators describe all the operators that exist in the Hilbert space of the symmetric states.

## Proof of universality

In this section, we prove that squeezing and coherent displacements provide universal control of symmetric states. The set of Hamiltonians $\{H_i\}$ is universal if any unitary can be constructed from their respective time evolution operators $U(t) = e^{-iH_i t}$. As was shown in [51], to prove universality, it is enough to show that the algebra derived from the set spans the entire Hilbert-space, i.e., that we can get any operator in the Hilbert-space by using linear combinations and commutators of $\{H_i\}$. Our Hilbert-space of matrices of size $(N + 1) \times (N + 1)$ is spanned by polynomials of the symmetric operators $S_x, S_y, S_z$, as shown in the Supplementary Information (SI). Hence, it is enough to show that we can construct any polynomial in $S_x, S_y, S_z$ using only the commutations and sums of Hamiltonians from the set $\{H_i\}$. In our case, this set is coherent rotations and squeezing, i.e., $\{H_i\} = \{S_x, S_y, S_x^2, S_y^2\}$.

It is insightful to compare the quantum systems described by symmetric states to those described by harmonic oscillators. In the case of harmonic oscillators, universality cannot be achieved using only coherent rotations and squeezing. The difference can be seen already in the commutation relation $[S_x, S_y] = iS_z$, which differs from that of the analogue operators $x$ and $p$ obeying $[x, p] = i\hbar$. For the symmetric operators, the commutation relation of two 1$^{st}$ order operators yields a 1$^{st}$ order operator, while for the harmonic oscillator operators, the commutation relation provides a scalar. Introducing the squeezing operators $S_x^2$ and $S_y^2$ allows us to reach the commutation relations $[S_x, S_y^2] = i(S_z S_y + S_y S_z) = S_x + iS_y S_z$. So, by subtracting $S_x$, we can get the operator $S_y S_z$. The same can be done with $[S_z, S_y^2]$ and $[S_y, S_x^2]$ to get $S_x S_y$ and $S_x S_z$. Now, we incorporate $S_y S_z$ in the commutation relation $[S_x, S_y S_z] = i(S_z^2 - S_y^2)$. Combining the other operators that we have achieved so far, we see that the algebra contains all polynomials of $S_x, S_y$ and $S_z$ of the 2$^{nd}$ power.

With all the 2$^{nd}$ power operators in hand, we can also construct any 3$^{rd}$ power operator by using commutation of all 2$^{nd}$ power operators, for example $[S_x^2, S_x S_y] = iS_x^2 S_z + iS_x S_z S_x$. With this polynomial, we can reach $S_x^3$ since $[[S_x^2, S_x S_y], S_y] = S_x^3$ + lower order terms. In this manner, we can build 4$^{th}$ power operators from 3$^{rd}$ power operators and so on as shown using an inductive proof in the SI.

## Optimization Protocol

To create arbitrary quantum states, we design and test an optimization protocol for engineering a sequence of coherent rotations $R(\theta, \hat{n}) = \exp(i\theta \vec{S} \cdot \hat{n})$ ($\vec{S} = S_x \hat{x} + S_y \hat{y} + S_z \hat{z}$) and squeezing operations $\exp(iS_x^2 \alpha)$ and $\exp(iS_y^2 \beta)$ (Fig. 1a). We demonstrate the protocol to yield several desired symmetric states, which translate to quantum light such as Schrödinger's cat and GKP states. We start from the ground state $|0\rangle$ and at each of $M$ steps, perform coherent rotation $R(\theta, \hat{n})$ followed by squeezing, which is a combination of $\exp(iS_x^2 \alpha)$ and $\exp(iS_y^2 \beta)$. We optimize the parameters $\hat{n}, \theta, \alpha, \beta$ of each step to maximize the fidelity between the final state $|\psi(\{\theta_i, n_i, \alpha_i, \beta_i\}_{i=1}^{M})\rangle$ and the target state $|\psi_{\text{target}}\rangle$ (Fig. 2). Throughout the paper we only use pure states, however our algorithm and code support noises and work with density matrices. We base our optimization protocol on a random search algorithm [52] applied over initial random guesses, and we then use the Nelder-Mead method [53] to find the local minimum for each initial guess (see the SI).

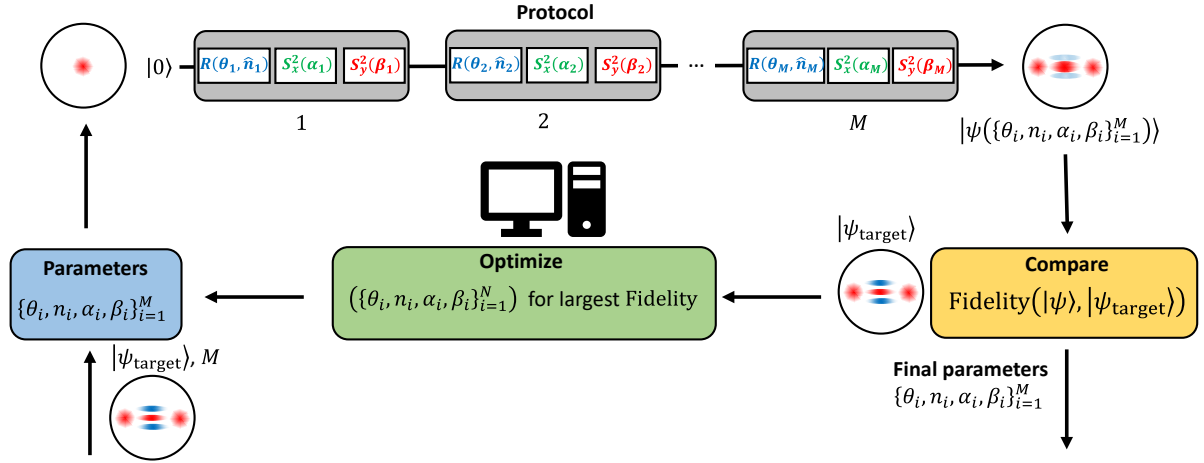

**Figure 2: Protocol for creating target symmetric states.** As an input, we set the target state $|\psi_{\text{target}}\rangle$ that we want to get, the desired number of sequences $M$, and the operations used in each sequence (as shown in Fig. 1). The optimization protocol finds the parameters used in each sequence to maximize the fidelity between the final state and the target state. Although we use pure states in the figure, the code works with density matrices.

Illustrating the strength of our approach, the protocol finds an eleven-steps sequence that produces a square GKP symmetric state with 98% fidelity. Fig. 3 presents the Wigner function [54] of the state on the Bloch sphere [48] at each of the steps.

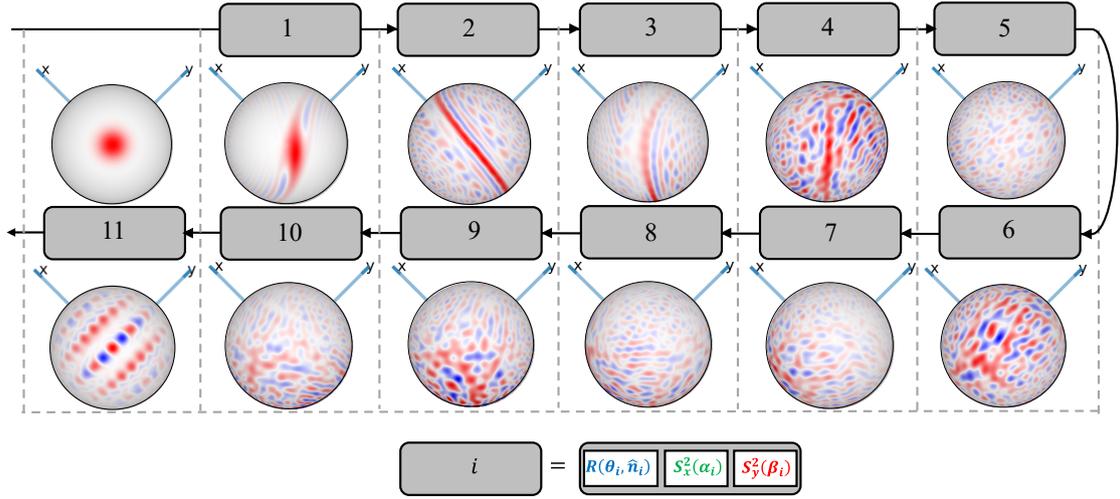

**Figure 3: Example sequence that creates a square GKP state.** Sequence for the creation of a GKP symmetric state with 10 dB squeezing using 11 steps displaying the symmetric state of the emitters after each step. Each $i^{th}$ step consists of coherent rotation and squeezing with $S_x^2$ and $S_y^2$ operators. The number of emitters in the simulation is 40, the fidelity of the final state is 98.37% compared to a square GKP symmetric state with 10 dB squeezing. For other parameters, see SI.

Our protocol also finds useful quantum states with high fidelity, such as two-legged and four-legged Schrödinger's cat states, as well as the hexagonal GKP state. Fig. 4 presents the fidelities, the number of steps, the Wigner-function of the resulted symmetric states on the sphere, and the corresponding emitted quantum states of the emitted light.

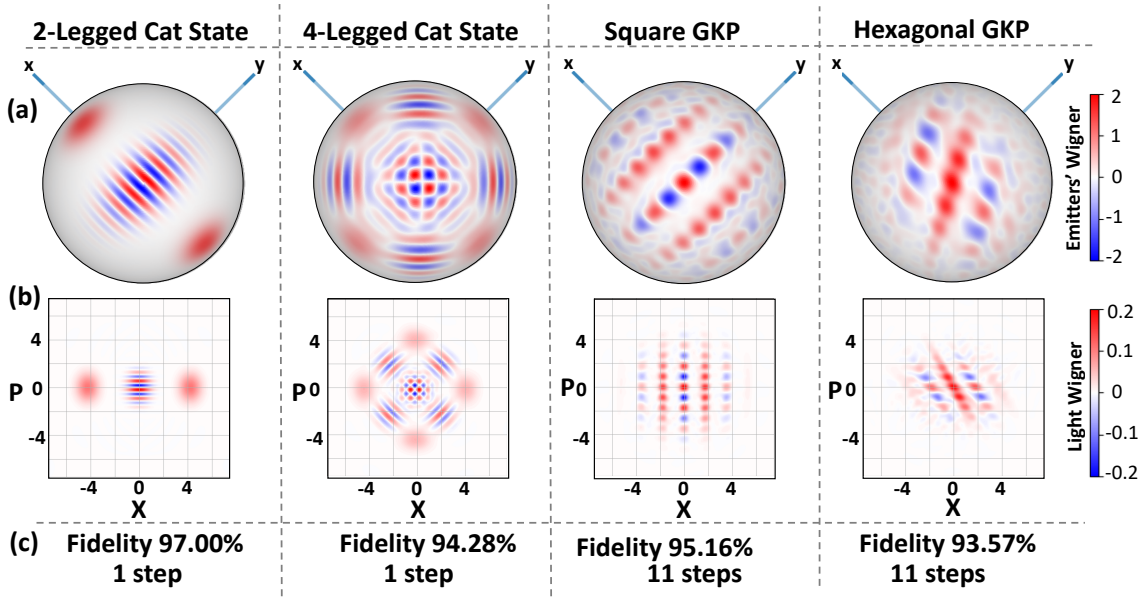

**Figure 4: Example states that can be created using sequences of squeezing and rotation operations.** **(a)** Two-legged and four-legged cat states, as well as square and hexagonal GKP states with 10 dB squeezing generated with our protocol (the third column corresponds to Fig. 3). **(b)** Corresponding Wigner functions of the emitted light states, assuming unity efficiency of coupling to a waveguide, using the model in [32]. **(c)** The fidelity between the target symmetric states and the ones achieved by the optimization protocol, denoting the number of steps. (The full list of parameters found by the optimization protocol can be found in SI).

**Discussion**

In this work, we showed that the simplest possible 1$^{st}$ and 2$^{nd}$ order operations (i.e., coherent rotations and spin squeezing) provide universal control over symmetric states of emitters. In contrast to standard two-qubit gate approaches, our method requires only a polynomial number of steps to create any symmetric state (see SI). We designed an optimization protocol that finds efficient sequences of such operations for the creation of arbitrary superpositions of symmetric states. Consequently, the creation of quantum states of emitters that are useful for quantum computation is possible without addressing the emitters individually (as in [55,56]).

States of emitters that are symmetric to permutations are especially interesting as they are the states that can be naturally mapped to arbitrary single-mode pulses of quantum light [32]. Using a full multimode theory [32], we analyze the spontaneous emission from the symmetric states created using our protocol. We find that the emission takes the form of single-mode traveling Schrödinger's cat and GKP states, which are desirable for quantum technologies. This result is in striking difference to most current methods that can create GKP states only in cavities in microwave frequencies (e.g., [40]).

Compared to previous works (e.g., [43]), which were so far limited to a linear regime where the number of emitters is greatly larger than the number of excitations, our approach explicitly utilizes the non-linearity that is intrinsic to the emitters at the high excitation regime. In addition, as we manipulate the emitters using rotations and squeezing, the transition into the coherent basis of the emitted light is natural, and thus the creation of Schrödinger's cat and GKP states takes a relatively small number of steps.

**Outlook**

Our protocols considered manipulation of the emitters only prior to their emission. We suspect that by manipulating the emitters during emission, using nonlinear operations like squeezing, as well as detuning of the emitter frequencies, exotic multimode states of light can be created efficiently, such as NOON, multimode squeezed states and cluster states.

The protocol presented in this work can be implemented in many modern quantum information processing platforms. For example, platforms of trapped ions [8] and superconducting circuits [6,7] could be ideal candidates as the symmetric operations considered in this work naturally appear in them. Furthermore, the rising field of waveguide quantum electrodynamics (QED) [57], along with the more mature field of cavity QED [11,58], are promising platforms for implementing our proposal, providing high collection and detection efficiency of the emitted light. We note that the decoherence mechanisms in each of these platforms can be analyzed using simple extensions of the protocol that we presented, and optimization under the effect of these mechanisms is straightforward.

Looking forward, we see the utilization of low-order nonlinearities such as spin squeezing in different platforms as an efficient method for universal control of emitters and for creating sources of both single and multimode quantum light. Our results emphasize that multipartite entangled states of emitters can be achieved without the need to control the emitters individually.

## Sources

GitHub repository: https://github.com/NGBigField/Superradiance

# Supplementary Information
# Universal control of symmetric states using spin squeezing

**Nir Gutman**[1†], **Alexey Gorlach**[1†], **Offek Tziperman**[1], **Ron Ruimy**[1] **and Ido Kaminer**[1*]

[1]*Technion – Israel Institute of Technology, Haifa 32000, Israel*

[†]*equal contributors,* [*]*kaminer@technion.ac.il,*

## Table of content

The supplementary information derives and supports the claims in the main text. We start by introducing the symmetric states of emitters and the operators that act on them. We continue by introducing the requirements for universal control and prove that a limited set of coherent rotations and squeezing operators is enough for universal control. We then show a bound on the number of steps needed to approximate any operation on symmetric states. Next, we present our optimization algorithm with which we found the exact rotations and squeezing that produce GKP and Schrödinger's cat states. We also summarize the parameters used for the results presented in the figures of the main text. Finally, we discuss the effects of the size of the system on the sequences.

**Symmetric operators and their properties**

As was done in [1,2], we start by presenting an ensemble of $N$ two-level identical emitters, each existing in the ground state $|g\rangle$, the exited state $|e\rangle$, or a superposition thereof. The lowering and raising Pauli operators of the specific system with index $i$ are $\sigma_i^- = |g\rangle\langle e|$ and $\sigma_i^+ = |e\rangle\langle g|$ accordingly that act on the system with the following relations $\sigma_i^+|g\rangle = |e\rangle$, $\sigma_i^-|e\rangle = |g\rangle$ and $\sigma_i^-|g\rangle = \sigma_i^+|e\rangle = 0$. Another important operator is the diagonal operator $\sigma_i^z = \frac{1}{2}(|g\rangle\langle g| - |e\rangle\langle e|)$. Each one of these operators acts exclusively on the $i^{th}$ site, with the commutation relations being:

$$[\sigma_i^z, \sigma_j^\pm] = \mp\delta_{ij}\sigma_i^\pm; \quad [\sigma_i^+, \sigma_j^-] = -2\delta_{ij}\sigma_i^z. \tag{S1}$$

If we have an ensemble of indistinguishable two-level emitters and we want to remain indistinguishable after an operation, this operation should be applied on the entire system in the same way on all two-level emitters. Only symmetric operations satisfy these conditions and can be constructed from the following basic operations:

$$S_\pm = \sum_{i=1}^{N} \sigma_i^\pm; \quad S_z = \sum_{i=1}^{N} \sigma_i^z. \tag{S2}$$

Since $S_+$ acting on the ground state $|gg\ldots g\rangle$ gives us $\sum_{i=1}^{N} \sigma_i^\pm |gg\ldots g\rangle = |eg\ldots g\rangle + |ge\ldots g\rangle + \cdots |gg\ldots e\rangle$, it is much more convenient to use the symmetric states:

$$\begin{cases} |N\rangle = |ee\ldots e\rangle \\ |N-1\rangle = 1/\sqrt{N}(|ge\ldots e\rangle + |eg\ldots e\rangle + \cdots |ee\ldots g\rangle) \\ \quad\quad\quad \ldots \\ |1\rangle = 1/\sqrt{N}(|eg\ldots g\rangle + |ge\ldots g\rangle + \cdots |gg\ldots e\rangle) \\ |0\rangle = |gg\ldots g\rangle \end{cases}.$$

Then, for example, $S_+|0\rangle = \sqrt{N}|1\rangle$. Generally, the lowering or raising operators acting on symmetric state $|m\rangle$ follow the relations:

$$S_+|m\rangle = \sqrt{(m+1)(N-m)}\,|m+1\rangle \triangleq d_+^m|m+1\rangle, \tag{S3a}$$

$$S_-|m\rangle = \sqrt{m(N-m+1)}\,|m-1\rangle \triangleq d_-^m|m+1\rangle, \tag{S3b}$$

$$S_z|m\rangle = (m - N/2)|m\rangle. \tag{S3c}$$

We also note that the commutation relations for the symmetric operators are the same as for Pauli matrices:

$$[S_z, S_\pm] = \mp S_\pm; \quad [S_+, S_-] = -2S_z. \tag{S4}$$

It is also common to define the operators $S_x = S_+ + S_-$ and $S_y = i(S_- - S_+)$ with commutation relation $[S_x, S_y] = iS_z$.

**Conditions for universality**

As was shown in [3], to prove that a set of Hamiltonians $\{H_i\}$ promises universal control over the Hilbert-space, it is enough to show that the algebra derived from the set spans the entire Hilbert-space, i.e., that we can get any operator in the Hilbert-space by using linear combinations and commutators of $\{H_i\}$.

First, we prove that the entire Hilbert-space of symmetric states of $N$ particles (i.e., the states $|0\rangle, |1\rangle, \ldots |N\rangle$) can be spanned by operating on the ground state $|0\rangle$ with unitary operations corresponding to the Hamiltonians that are the polynomials of $S_x$ and $S_y$ up to order

$N$. A general symmetric state can be written in the symmetric basis $|\psi\rangle = \sum_{n=0}^{N} a_n |n\rangle$. Each term $a_n|n\rangle$ in the sum can be created from the ground state $|0\rangle$ by the operator $\frac{a_n}{c_n} S_+^n$, since $S_+^n |0\rangle = c_n |n\rangle$, where $c_n = \sqrt{\frac{n! N!}{(N-n)!}}$. Then $|\psi\rangle = \sum_{n=0}^{N} a_n |n\rangle = \sum_{n=0}^{N} \frac{a_n}{c_n} S_+^n |0\rangle$.

Now we notice that we can build the state $|\psi\rangle = \sum_{n=0}^{N} \frac{a_n}{c_n} S_+^n |0\rangle$ by applying infinitesimal unitary operations $e^{\alpha_1 S_+ - \alpha_1^* S_-}$ on $|0\rangle$, where $\alpha_1$ is a complex number such that $|\alpha_1| \ll 1$. Applying this unitary operation, we get:

$$e^{\alpha_1 S_+ - \alpha_1^* S_-} |0\rangle \approx |0\rangle + \alpha_1 S_+ |0\rangle = |0\rangle + \alpha_1 c_1 |1\rangle.$$

Then applying the same operator $M_1$ times yields:

$$\left(e^{\alpha_1 S_+ - \alpha_1^* S_-}\right)^{M_1} |0\rangle \approx |0\rangle + \alpha_1 M_1 c_1 |1\rangle.$$

We can choose $\alpha_1 M_1 = a_1/(a_0 c_1)$ and get $\left(e^{\alpha_1 S_+ - \alpha_1^* S_-}\right)^{M_1} |0\rangle \approx |0\rangle + \frac{a_1}{a_0} |1\rangle$. In this way, we were able to construct the coefficient of $|1\rangle$ in the wavefunction $|\psi\rangle$. To construct the coefficient of $|n\rangle$, we apply $M_n$ times the following unitary operator:

$$\left(e^{\alpha_n S_+^n - \alpha_n^* S_-^n}\right)^{M_n} |0\rangle \approx |0\rangle + \alpha_n M_n c_n |n\rangle.$$

Then we can choose $\alpha_n M_n = a_n/(a_0 c_n)$. Thus, by following this construction, we can build an arbitrary symmetric state $|\psi\rangle$:

$$\left(e^{\alpha_N S_+^N - \alpha_N^* S_-^N}\right)^{M_N} \ldots \left(e^{\alpha_2 S_+^2 - \alpha_2^* S_-^2}\right)^{M_2} \left(e^{\alpha_1 S_+ - \alpha_1^* S_-}\right)^{M_1} |0\rangle \xrightarrow[\{a_i\}_{i=1}^N \to 0]{} |0\rangle + \frac{a_1}{a_0} |1\rangle + \cdots \frac{a_n}{a_0} |N\rangle \propto |\psi\rangle.$$

This set of unitary operators corresponds to the Hamiltonians that are polynomials in $S_\pm = \frac{S_x \pm i S_y}{2}$ up to the $N^{\text{th}}$ order. Consequently, if we can construct any Hamiltonian that is an arbitrary polynomial in $S_x$ and $S_y$, then we can generate any pure symmetric state.

**Proof of universality**

Let us now discuss harmonic oscillators and the symmetric states of emitters. Both models contain coherent rotations and squeezing operations. However, as we will show below, only in two-level identical emitters squeezing and coherent rotations are enough for the universality, while for the harmonic operator these operations can only create Gaussian states.

Coherent rotations and squeezing do not constitute universal control over the states of the harmonic oscillator: Coherent rotations are the operators $x$ and $p$ and any linear combination thereof in a quadrature. Together with the 2$^{\text{nd}}$ "squeezing" operators $x^2$, $p^2$ and

$s = \frac{xp+px}{2}$, we get the set $\{x, p, x^2, p^2, s\}$. As was shown in [4], we start by writing some commutation relation. Two 1st order operators give a 2nd order commutation relation $[x, p] = i\hbar$ that yields a 0th order scalar. The 2nd order operator $s$ with both 1st order operators $x$ and $p$ yields the 3rd order commutation relations $[s, p] = -p$ and $[s, x] = -ix$, both yielding 1st order operators. Thus, we conclude that we can construct Hamiltonians that are quadratic in $x$ and $p$ and not higher. We do not span the entire Hilbert space and thus cannot achieve universal control.

Coherent rotations and squeezing constitute universal control over the symmetric states of emitters: In contrast to the previous system, here the coherent rotations are operators $S_x = S_+ + S_-$ and $S_y = i(S_- - S_+)$ with commutation relation $[S_x, S_y] = iS_z$. Already we see a stark contrast from the previous case. The commutation relation of two 1st order operators yields a 1st order operator. Introducing squeezing $S_x^2$ and $S_y^2$ allows us to reach the commutation relations $[S_x, S_y^2] = i(S_z S_y + S_y S_z) = S_x + iS_y S_z$. So, by subtracting $S_x$ we can get the operator $S_y S_z$. The same can be done with $[S_z, S_y^2]$ and $[S_y, S_x^2]$ to get $S_x S_y$ and $S_x S_z$. Now, since we already posses $S_y S_z$ and can incorporate it in the commutation relation, we get $[S_x, S_y S_z] = i(S_z^2 - S_y^2)$. Combining the other operators achieved so far, we see that the algebra contains all polynomials of $S_x, S_y$ and $S_z$ of the 2nd power.

With all the 2nd power operators in hand, we can also construct any 3rd power operator, and with them all 4th power operators and so on using the following inductive proof:
Assume that we can already reach $S_x, S_y$ and $S_z$ to some powers $n, m$ and $k$ respectively, we can then use commutation relations to get:

$$[S_x, S_y^m] = i \sum_{i=1}^{m} S_y^{i-1} S_z S_y^{m-i} = iS_y^{m-1} S_z + \text{lower orders}$$

$$[S_x^2, S_y^m] = S_x[S_x, S_y^m] + [S_x, S_y^m]S_x = i\sum_{i=1}^{m} S_x S_y^{i-1} S_z S_y^{m-i} + i\sum_{i=1}^{m} S_y^{i-1} S_z S_y^{m-i} S_x$$
$$= iS_x S_y^{m-1} S_z + \text{lower orders}$$

Thus:
$$[S_x^2, S_x^n S_y^m S_z^k] = S_x^n[S_x^2, S_y^m]S_z^k + S_x^n S_y^m[S_x^2, S_z^k]$$
$$= iS_x^{n+1} S_y^{m-1} S_z^{k+1} + iS_x^{n+1} S_y^{m+1} S_z^{k-1} + \text{lower orders} \quad \text{(S5a)}$$
$$[S_y^2, S_x^n S_y^m S_z^k] = iS_x^{n-1} S_y^{m+1} S_z^{k+1} + iS_x^{n+1} S_y^{m+1} S_z^{k-1} + \text{lower orders} \quad \text{(S5b)}$$
$$[S_z^2, S_x^n S_y^m S_z^k] = iS_x^{n-1} S_y^{m+1} S_z^{k+1} + iS_x^{n+1} S_y^{m-1} S_z^{k+1} + \text{lower orders} \quad \text{(S5c)}$$

Combining the formulas yields $[S_x^2, S_x^n S_y^m S_z^k] + [S_y^2, S_x^n S_y^m S_z^k] - [S_z^2, S_x^n S_y^m S_z^k] = 2i S_x^{n+1} S_y^{m+1} S_z^{k-1}$ + lower orders, effectively raising the power of $S_x$ and $S_y$ by 1 while only lowering the power of $S_z$ by 1, thus achieving set of operators higher in overall polynomial order than what we have started with. With different linear combinations, we can also get:

$$S_x^{n+1} S_y^{m+1} S_z^{k-1}, \quad S_x^{n+1} S_y^{m-1} S_z^{k+1}, \quad S_x^{n-1} S_y^{m+1} S_z^{k+1}$$

And hence any polynomial of $S_x, S_y$ and $S_z$ is reachable with enough mixing of the commutation relations and linear combination of them. The power to reach any polynomial promises universal control.

**Complexity**

With the proof that any unitary acting on symmetric states can be created using the simple set of coherent-rotations and squeezings $\{S_x, S_y, S_z, S_x^2, S_y^2\}$, a sensible question would be to ask how many such operations are needed to construct an arbitrary unitary operator. This, of course, will vary depending on the desired operator, so a better query will be to find a cap on the number of operations needed. In this section, we prove the needed number of operations to be polynomial in the number of qubits $N$, and in fact, show it to be $O(N^4)$.

Given a set of Hamiltonians $\{S_x, S_y, S_z, S_x^2, S_y^2\}$, we can construct an arbitrary product of the following operators $e^{-i\alpha S_{x,y,z}}, e^{-i\beta S_{x,y}^2}$ for arbitrary $\alpha$ and $\beta$. To effectively construct the unitary operator that is generated from the addition of Hamiltonians $A$ and $B$, we need $2k$ such operators, where $k$ is an integer with which greater accuracies can be achieved:

$$e^{-i(A+B)t} = \left(e^{-\frac{iAt}{k}} e^{-\frac{iBt}{k}}\right)^k + O\left(\frac{1}{k}\right).$$

To build the unitary generated from the commutation of $A$ and $B$, we need $4k$ operators:

$$e^{-i[B,A]t} = \left(e^{-iA\sqrt{\frac{t}{k}}} e^{-iB\sqrt{\frac{t}{k}}} e^{iA\sqrt{\frac{t}{k}}} e^{iB\sqrt{\frac{t}{k}}}\right)^k + O\left(\frac{1}{k}\right).$$

Now let us calculate how many operations we need to construct a desired unitary operator $U_i = e^{-iH_i t}$. We can examine what happens when this unitary acts on the ground state and notice that we will get some symmetric state $|\psi_i\rangle = e^{-iH_i t}|0\rangle \approx (1 - iH_i t)|0\rangle$. Since $|\psi_i\rangle$ can be decomposed into the symmetric basis as $|\psi_i\rangle = \sum_{n=0}^{N} a_n |n\rangle = \sum_{n=0}^{N} a_n S_+^n |0\rangle$, we find that $H_i$ can be decomposed into powers of $S_+$ operator up to $S_+^N$ (the higher powers of $S_+$ all give zero). This puts a cap of $N$ on the degree of the polynomial of $S_x, S_y, S_z$ in each Hamiltonian $H_i$, i.e.:

$$H_i = \sum_{n=0}^{N} \sum_{k=0}^{n} \sum_{m=0}^{n-k} c_{nmk} S_x^{n-k-m} S_y^k S_z^m \tag{S6}$$

For each degree $n$ from 0 to $N$ in (S6), the number of operators with this degree is $\sum_{k=0}^{n} \sum_{m=0}^{n-k} 1 = \frac{1}{2}(n+1)(n+2) \in O(n^2)$.

We showed universality using a proof by induction, where a polynomial of $S_x, S_y, S_z$ with degree $n$ can be constructed by using $O(n)$ commutation relations, i.e., with $O(n \cdot k)$ unitary operations, using polynomials of lower degrees $(n-1, n-2, \ldots, 2, 1)$.

Meaning that to get all polynomial of degree $n$, we need only $O(n^3 \cdot k^3)$ operators of lower degrees, and thus $O(N^4 \cdot k^3)$ operators to get all polynomials up to degree $N$. Hence, the overall number of operators from the set $\{S_x, S_y, S_z, S_x^2, S_y^2\}$ needed to construct any unitary in the Hilbert space of symmetric states of $N$ qubits is $O(N^4 \cdot k^3)$. Thus, we proved that any unitary acting in the basis of symmetric states can be achieved by applying $O(N^4 \cdot k^3)$ operations from the set of symmetric operators $\{S_x, S_y, S_z, S_x^2, S_y^2\}$. To get good precision one needs a large enough $k$, however the number of operations even for large $k$ still scales as $N^4$. In practice, many interesting states can be achieved with even less operations, as we have shown in the main text.

**Optimization protocol**

The optimization algorithm's purpose is to find the optimal parameters for the sequence that results in a state with maximal fidelity with respect to the target state: $\text{Fidelity}\left(\rho_{\text{target}}, \rho(\vec{\theta})\right)$, where the set of parameters $\vec{\theta}$ control the different unitary operations at each time step towards the final state. The parameter space is filled with local minima, that greatly hinders most gradient-descent based optimization methods.

To overcome this difficulty, we introduce two random processes into our method. Firstly, the process of optimization begins with the generation of many independent random guesses as the initial starting parameters. Secondly, for each such set of initial parameters, we repetitively start and optimization instance of the Nelder-Mead method [5], where some of the parameters are locked and do not participate in the optimization method. This allows us to sometimes skip over some local minima, in hope to eventually converge into the global minimum.

The Nelder-Mead method was chosen because it was the only readily available method which proved to find local minima in our system while incorporating the constraints that we deemed necessary (e.g., limiting the angles of rotations to the range $[-\pi, \pi]$) and does not require knowledge of the exact derivatives. We believe that if an exact derivative can be computed for any parameter in the sequence of pulses, a better optimization method could have been chosen.

Each rotation pulse was parametrized by the amount of clockwise rotation (can be also negative to represent a counterclockwise rotation) around each of the primary axes $\hat{x}, \hat{y}$ and $\hat{z}$. This method replaces an equally valid alternative method that represents any rotation as some angle of rotation around some axis of rotation (as promised by Euler's rotation theorem).

The number of required pulses $N$ is a hyperparameter, unknown prior to finding the best solution. To find it, we always started with only two pulses, exhaustively tried many initial guesses until it seems that a better fidelity cannot be achieved, and then only then increased $N$ by 1. When adding another step of rotation and squeezing to the sequence (thus, increasing N), we initialize all its parameters to 0. Which acts the same as adding the identity operator. This allows us to add the new step in the middle (or even beginning) of the sequence, giving the algorithm a richer parameter space to work with, while not losing any progress gained thus far.

For each squeezing pulse we have two parameters, one for each direction ($\hat{x}$ and $\hat{y}$) and three parameters for its preceding rotation – overall five free parameters for each pulse plus three parameters for a final rotation, giving:

$$\text{Number of Parameters} = 5 \cdot N + 3. \tag{S6}$$

For typical values of $N$ in the main text, this means having above ~30 parameters, and thus most derivative-less methods will struggle to converge on a global minimum in a parameter-space filled with local minima. This is the reason why we introduced the second random process of freezing part of the parameters while letting the others take part in the optimization process. We have found that a good number of free parameters is ~20. This is also the key to allowing new pulses to be added in the middle of a sequence: Add a new pulse with all zero initial parameters, freeze most parameters besides the new ones, and start a new optimization instance.

**Parameters used in the numerical simulations**

Here we write explicitly the parameters that are needed to replicate the results shown in the main paper. The rotation parameters at each step are: $\hat{n}_i$ for the rotation axis and $\theta_i$ for the degree of rotation (in radians). The squeezing in directions x and y are $\alpha_i$ and $\beta_i$ respectively. In the cat codes, we write $|\gamma\rangle = e^{-\frac{1}{2}|\gamma|^2} \sum_{n=0}^{\infty} \frac{\gamma^n}{\sqrt{n!}} |n\rangle$ for a coherent state written in the symmetric basis, with a complex parameter $\gamma$.

2-legged Schrödinger's cat:

Number of emitters = 40

The state is $|2cat\rangle = |\gamma\rangle - i|-\gamma\rangle$ ×(normalization factor), where $\gamma = 3$.

| step # | $\hat{n}$ | | | $\theta$ | $\alpha$ | $\beta$ |
|---|---|---|---|---|---|---|
| 1 | 0.44683 | 0.21351 | 0.86876 | 3.73127 | 1.57143 | 1.57017 |
| end | 0.00000 | 0.00000 | 1.00000 | 2.35948 | | |

4-legged Schrödinger's cat:

Number of emitters = 40

The state is $|4cat\rangle = |\gamma\rangle + e^{i\phi}|+i\gamma\rangle + e^{i2\phi}|-\gamma\rangle + e^{i3\phi}|-i\gamma\rangle$ ×(normalization factor), where $\gamma = 3$ and $\phi = \pi/4$.

| step # | $\hat{n}$ | | | $\theta$ | $\alpha$ | $\beta$ |
|---|---|---|---|---|---|---|
| 1 | 0.220058 | 0.789986 | -0.57227 | 4.061466 | 0.003675 | -0.7836 |
| end | 0.614706 | 0.530748 | -0.58347 | 4.24025 | | |

Hexagonal GKP:

Number of emitters = 40. Squeezed by 10 dB.

| step # | $\hat{n}$ | | | $\theta$ | $\alpha$ | $\beta$ |
|---|---|---|---|---|---|---|
| 1 | 0.58422 | -0.70216 | 0.40701 | 3.73793 | -0.01883 | 0.10958 |
| 2 | 0.12109 | 0.76554 | 0.63189 | 3.51680 | 0.56324 | 0.06183 |
| 3 | -0.99588 | 0.03711 | 0.08277 | 3.16463 | -0.46139 | 0.24754 |
| 4 | 0.55086 | 0.73301 | -0.39905 | 0.81578 | 0.01654 | 0.05930 |
| 5 | 0.63164 | 0.09674 | 0.76920 | 0.46043 | 0.00749 | 0.01291 |
| 6 | 0.17495 | 0.05641 | -0.98296 | 1.17352 | 0.01198 | 0.02608 |
| 7 | 0.69954 | -0.23653 | -0.67431 | 3.11339 | 3.15159 | -0.01382 |
| 8 | 0.70298 | -0.64758 | -0.29404 | 0.99739 | -1.37202 | 0.03684 |
| 9 | 0.89993 | 0.00384 | 0.43603 | 0.77684 | 0.52291 | 0.25252 |
| 10 | 0.89155 | 0.38970 | -0.23081 | 0.41266 | 3.07564 | 0.79775 |
| 11 | 0.05450 | 0.33542 | 0.94049 | 1.54108 | 0.24451 | 1.06955 |
| end | 0.92003 | 0.38822 | -0.05321 | 3.42003 | | |

Square GKP:

Number of emitters = 40. Squeezed by 10 dB.

| step # | $\hat{n}$ | | | $\theta$ | $\alpha$ | $\beta$ |
|---|---|---|---|---|---|---|
| 1 | -0.00355 | 0.307337 | -0.95159 | 0.396325 | 0.024936 | 0.227007 |
| 2 | -0.01773 | 0.230252 | -0.97297 | 1.090371 | 0.002457 | 0.21566 |
| 3 | 0.200697 | 0.978824 | -0.0403 | 0.159439 | -0.00392 | -0.06637 |
| 4 | 0.806244 | -0.36774 | 0.463401 | 3.038831 | -0.04836 | -0.29011 |
| 5 | -0.63756 | 0.630441 | -0.44278 | 3.500829 | -1.10497 | -0.24593 |
| 6 | 0.020905 | -0.92403 | -0.38174 | 2.812456 | 0.028772 | -0.11394 |
| 7 | -0.76349 | -0.62554 | 0.160584 | 0.350578 | 0.40291 | 0.167819 |
| 8 | 0.96056 | -0.14113 | -0.2396 | 1.718151 | -0.00775 | 0.103055 |
| 9 | 0.854513 | 0.223536 | 0.46887 | 0.770651 | 0.044993 | -0.06539 |
| 10 | 0.218843 | -0.92957 | 0.296648 | 0.256411 | -0.08197 | 0.049369 |
| 11 | 0.321551 | -0.6716 | -0.6675 | 0.506597 | -0.30929 | 0.992603 |
| end | -0.00781 | -0.66735 | 0.744704 | 1.456135 | | |

Some parameters appear in higher precision in the code.

**Different number of the emitters**

It is important to note that the generation of different quantum states is strongly connected to the fact that the Wigner function of the emitters is defined on the Bloch sphere. While there is an infinite number of states of the harmonic oscillator, there are only $N + 1$ symmetric states that arise from $N$ indistinguishable emitters. This fact is linked to the observation that the displacement operators on states of the harmonic oscillator can move coherent states to any position $|\alpha\rangle$ on the quadrature, infinitely far away from $|0\rangle$, while coherent rotations on symmetric states are limited by the size of the Bloch-sphere they exist on.

Coherent rotations that rotate coherent states close to the bottom of the sphere are not much different from those imagined on an infinite plane, but using the top of the sphere is unique to a system of emitters and does not exist in photonic systems. Thus, the sequence of unitary transformation leading to the target symmetric state is strongly dependent on the number of emitters in the system (i.e., the size of the sphere) (Fig. S1). Photonic systems, in comparison, has no such parameter.

We can deduct two key statements from this observation: Firstly, the limited number of symmetric states plays a key role in our ability to achieve universal control from such a small set of operators. Secondly, if during a sequence the Wigner function on the Bloch-sphere has values too far above the equator (close to the top of the sphere), we expect this sequence to get entirely different results for a different system size. Figure S1 shows how the same sequence can result in entirely different final states when the size of the system varies.

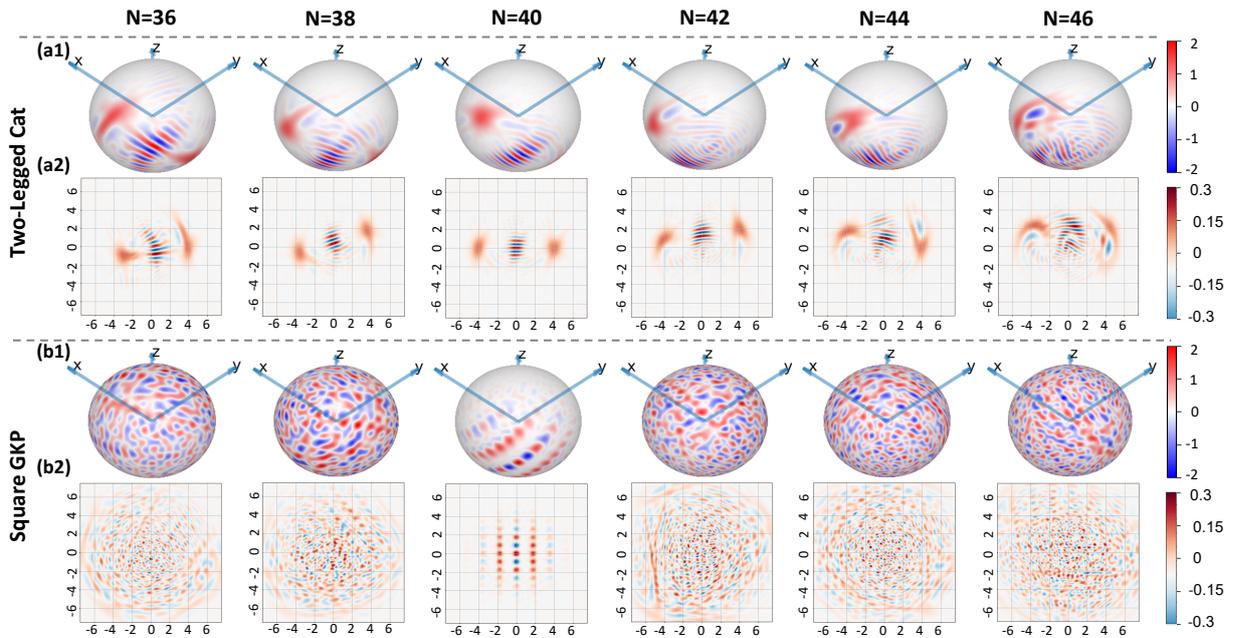

**Fig. S1: The dependence of the sequence on the number of emitters:** Final states using the sequences that we have learned for a system with 40 emitters. Matter-states on the Bloch-sphere **(a1)** and the projections onto the plane **(a2)** for a sequence optimized to create a 40-emitters two-legged cat, used on different systems with different sizes $N$. **(b1-2)** The same results for a sequence optimized for square-GKP. The Wigner-function of the matter-state during the sequence optimized for the two-legged cat stays close to the lower half of the sphere, and thus the size of the system does not affect the result as much as for the sequence optimized for a square-GKP that does involve all of the sphere.

## Sources

GitHub repository: https://github.com/NGBigField/Superradiance